# Dynamic read mapping and online consensus calling for better variant detection

Karel Břinda[1,*], Valentina Boeva[2,3], Gregory Kucherov[1]

**1** Université Paris-Est, LIGM, CNRS, 77454 Marne-la-Vallée, France, **2** Institut Curie – Centre de Recherche, PSL Research University, Mines Paris Tech, INSERM U900, 75005 Paris, France, **3** Institut Cochin, INSERM U1016, CNRS UMR 8104, Université Paris Descartes UMR-S1016, 75014 Paris, France

\* Corresponding author
E-mail: `karel.brinda@univ-mlv.fr` (KB)

## Abstract

Variant detection from high-throughput sequencing data is an essential step in identification of alleles involved in complex diseases and cancer. To deal with these massive data, elaborated sequence analysis pipelines are employed. A core component of such pipelines is a read mapping module whose accuracy strongly affects the quality of resulting variant calls.

We propose a *dynamic read mapping* approach that significantly improves read alignment accuracy. The general idea of dynamic mapping is to continuously update the reference sequence on the basis of previously computed read alignments. Even though this concept already appeared in the literature, we believe that our work provides the first comprehensive analysis of this approach.

To evaluate the benefit of dynamic mapping, we developed a software pipeline (`http://github.com/karel-brinda/dymas`) that mimics different dynamic mapping scenarios. The pipeline was applied to compare dynamic mapping with the conventional static mapping and, on the other hand, with the so-called *iterative referencing* – a computationally expensive procedure computing an optimal modification of the reference that maximizes the overall quality of all alignments. We conclude that in all alternatives, dynamic mapping results in a much better accuracy than static mapping, approaching the accuracy of iterative referencing.

To correct the reference sequence in the course of dynamic mapping, we developed an online consensus caller named Ococo (`http://github.com/karel-brinda/ococo`). Ococo is the first consensus caller capable to process input reads in the online fashion.

Finally, we provide conclusions about the feasibility of dynamic mapping and discuss main obstacles that have to be overcome to implement it. We also review a wide range of possible applications of dynamic mapping with a special emphasis on variant detection.

## Introduction

High-throughput sequencing technologies (Next-Generation Sequencing, commonly abbreviated NGS) made available terabases of DNA sequence data coming from numerous de-novo and resequencing experiments. A central computational problem in NGS data analysis is inferring the original DNA sequence of an individual from a large set of NGS reads, i.e., short sequence fragments generated by a sequencing machine. This task is usually solved by mapping reads to a high-quality reference sequence or by a computationally expensive genome assembly algorithm.

In this paper, we study genotyping by sequencing when a reference sequence is available. A typical pipeline for inferring the genotype of a sequenced individual consists of the following steps (see, e.g, [1,2]). Reads obtained from a sequencer are mapped to the reference sequence and low quality alignments are discarded.

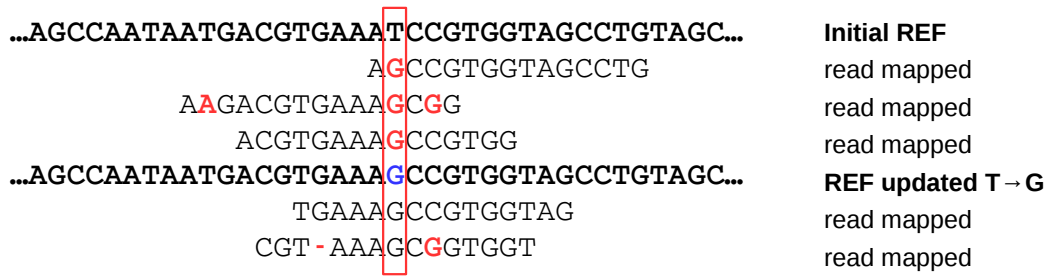

**Figure 1. Demonstration of idea of dynamic mapping.** In dynamic mapping, reference sequence is updated according to already mapped reads. In this example, aligned reads suggest that the base at the highlighted position is G rather than T, and the reference is corrected as soon as G is supported by sufficiently many reads. To guide the updates, a dynamic mapper must be equipped with an online consensus caller distinguishing SNPs and sequencing errors. Disagreements between aligned reads and the reference (sequencing errors, SNPs and indels) are highlighted in red and updates of the reference in blue.

Aligned reads are further sorted and possibly locally re-assembled or re-aligned around indels in order to better distinguish SNPs from indels. Systematic biases in base qualities can be further removed from the reads in a base quality recalibration step. Then, variants are called (either for each individual separately or jointly for a group of them) and genotype likelihoods inferred. At the final step, variant qualities are recalibrated, and variants filtered and annotated.

In such a pipeline, read mapping is a crucial step affecting overall results. The ultimate goal of a read mapper [3, 4] is to align NGS reads or their parts to their "true origin" in the reference genome, inferring their positions and occurred variants. To achieve this goal, each read is aligned to the most similar region(s) in the reference determined by the alignment scores.

There are several factors that make mapping a challenging task in practice. Genome sequences are highly repetitive and often contain several regions equally similar to a read. Employing a sequencing technology producing long or paired-end reads can alleviate but not completely eliminate this difficulty. Reliability of mapping strongly depends on how much the sequenced genome differs from the reference. Close individuals (such as those of the same species) usually differ only weakly and highly variable regions tend to be rare. For distant individuals, evolutionary events (such as genomic rearrangements or gene duplications) may cause more substantial differences between the two genomes making it impossible to deduce the true origin of some reads using similarity search. Sequencing errors constitute another major obstacle. Their rate strongly depends on the employed technology – from Illumina HiSeq producing reads of 100bp with about 1% of errors to PacBio SMRT and Oxford Nanopore producing reads of up to several tens of thousand bp with about 15% of errors – and on the experimental conditions. Furthermore, some reads may come from regions absent in the reference sequence or may result from contamination by DNA from other genomes. Such reads should be detected and annotated by the mapper, i.e. marked as unaligned or as aligned with low quality. Finally, the sheer volume of input data makes of mapping a computationally demanding task constraining the choice of the underlying algorithm. For instance, whole genome sequencing projects of the human genome of 3 Gbp often use more than $30\times$ coverage.

As a consequence of the above, practical read mappers must represent a fine-tuned trade-off between memory requirements, speed and sensitivity, and incorporate specific characteristics of target sequencing technologies. The software must be highly optimized, parallelized, and well-debugged. Today, most common practical mappers include BWA-MEM [5], BOWTIE 2 [6], GEM [7], LAST [8], and NOVOALIGN. These mappers align reads coming in a stream, in the online fashion, against a fixed reference sequence. Since the reference is unchanging, we call such an approach *static mapping*.

Despite a great deal of work on mapping techniques and a large number of existing mappers (see, e.g., lists on `http://bit.ly/1Y1eNas` or `http://bit.ly/1sUwsa3`), the mapping process remains error prone and often produces wrong placement of reads, partial mappings, or incorrectly considers reads as non-mappable. Furthermore, another inherent drawback of static mapping is the bias introduced by alleles which differ between the reference and sequenced genomes as alleles present in the reference tend to be favored in the alignment of reads. This may seriously affect the results of the analysis [9].

In the course of mapping, read alignments computed so far provide a useful information about allelic differences between sequenced and reference sequences, which can be a helpful guide for adjusting future mappings, or can even call for remapping previously mapped reads. In this paper, we study the *dynamic mapping* approach when the reference sequence is continuously updated based on already computed alignments, and show that this can significantly improve the quality of mapping. Fig. 1 provides an illustration to this idea. The general goal of our work is to analyse the pros and cons of dynamic mapping and to provide accurate quantitative estimates supporting our analysis.

Design and implementation of dynamic mapping constitute a difficult task. One major obstacle is that the underlying index data structure for the reference sequence must support dynamic updates. There are two main types of indexing structures used by modern mappers. The first one relies on *full-text indexes*, in particular on BWT-index [10], and is used, among others, by the BWA family [5,11,12], GEM [7], or Bowtie 2 [6]. Another group is based on *hash tables* and includes such mappers as SHRiMP2 [13], PerM [14], NovoAlign, BFast [15], StamPy [16], and others. Generally speaking, hash table-based structures require a much larger memory space and, therefore, projects dealing with large data (such as whole eukaryotic genomes) usually use mappers of the first group. On the other hand, BWT-index hardly lends itself to dynamic updates. One attempt to implement dynamic updates of BWT-index was made in [17–19], another approach was studied in [20]. In both cases, provided implementations are only experimental and cannot be considered as ready-to-use tools for large-scale dynamic mapping. A dynamic $k$-mer-based index has very recently been proposed in [21].

Another important component of dynamic mapping is online consensus calling which guides the updates of the reference. An online consensus caller must support quick updates of the reference after new read alignments have been computed. This requires keeping, in a compact form, a genome-wide statistics of possible variants, such as a truncated pileup information. Finally, a dynamic mapper may have to support remapping of already mapped reads.

Dynamic mapping has been little studied so far and only two partial experimental solutions with naive mapping algorithms are available [22,23]. Program DynMap [22] implements a data structure specifically designed for mapping reads to a dynamically updated sequence. Unfortunately, space complexity of DynMap is linear in the number of reads which makes it hardly applicable to contemporary sequencing technologies. Manuscript [23] (apparently unpublished) is the first attempt to systematically analyse the benefit and the overhead of dynamic mapping using a simple dedicated tool (https://github.com/jpritt/fm-update). With an index based on dynamic suffix array [19], the accuracy of dynamic mapping has been compared to iterative referencing, depending on the mutation rate, read length, number of reads and sequencing error rate. The author of [23] also considered the problem of statistics reduction and suggested to use coverage vectors in order to take only first $k$ reads for each base of the reference.

In this paper, we present a comprehensive and systematic study of dynamic mapping. We first focus on the problem of online consensus calling which is a component required for dynamic mapping. We further present the first online consensus caller, named Ococo. We present a software pipeline for simulating various alternatives of dynamic mapping. We then provide a comparative analysis of different dynamic mapping scenarios on several datasets simulated from bacterial genomes, using the previously developed RNFtools framework [24]. Obtained results show that dynamic mapping can provide a significant improvement of the accuracy of read alignment compared to traditional mapping approaches. The pipeline software and the results of our experiments are available from http://github.com/karel-brinda/dymas.

# Materials and Methods

## Online consensus calling

A dynamic mapper has to collect statistics about previously mapped reads. When this statistics accumulates a sufficient evidence of a difference at a certain locus between sequenced and reference genomes, an update of the reference is reported that triggers a corresponding update of the index. This process is the same as the usual consensus calling except that it has to be performed in the online (streamed) fashion. Later in the Discussion section we will provide further remarks on the biological relevance of the consensus and on the effect of ploidy.

| Incoming base | A-counter bin | dec | C-counter bin | dec | G-counter bin | dec | T-counter bin | dec | Sum | |
|---|---|---|---|---|---|---|---|---|---|---|
|  | 010 | 2 | 110 | 6 | 001 | 1 | 010 | 2 | 11 | Initial state |
| C | 010 | 2 | 111 | 7 | 001 | 1 | 010 | 2 | 12 | C-counter incremented |
| T | 010 | 2 | 111 | 7 | 001 | 1 | 011 | 3 | 13 | T-counter incremented |
| C | 001 | 1 | 011 | 3 | 000 | 0 | 001 | 1 | 5 | All counters bit-shifted to right |
|  | 001 | 1 | 100 | 4 | 000 | 0 | 001 | 1 | 6 | C-counter incremented |

**Figure 2. Internal statistics of an online consensus caller.** This example shows how 3-bit counters keep reduced truncated pileup information for a single position. The counters with initial values $(2, 6, 1, 2)$ are incremented after nucleotides C, T, C have been received for this position. After receiving C and T, the corresponding counters have been simply incremented. Then, after receiving C, all counters are bit-shifted before the C-counter is incremented.

In this work, we consider consensus calling as SNP calling using reads coming from a single individual. Note that throughout this paper single nucleotide insertions and deletions are considered as SNPs. SNP calling is a widely studied problem (see, e.g., [1, 25–34]) but to the best of our knowledge, all published solutions are offline, i.e, all reads have to be aligned and sorted prior to the calling step. Variants are inferred from differences between aligned reads and the reference and are usually stored in VCF format [35] together with confidence scores. They are then filtered in order to keep only those which are reliable enough, in particular those which do not correspond to sequencing errors. After that, they are incorporated into the original reference to obtain the consensus.

In this section, we briefly describe OCOCO, an online consensus caller that we developed within this work.

### Consensus calling algorithm

Regular offline variant callers usually operate by sliding a small window through the genome and keeping in memory information (read alignments, mapping qualities, base qualities) collected within individual windows. After variants are detected, their significance scores are computed using the underlying statistical model.

In online consensus calling, reads come from a stream and can map to random positions, therefore our "window" must be the entire genome. As a consequence, we can afford storing only a limited information about processed read alignments. OCOCO only deals with single-nucleotide replacement updates, other types of updates will be considered in Discussion. OCOCO supports two distinct working modes of an online consensus caller. In the *real-time mode*, suitable for dynamic mapping, updates are computed and incorporated into the reference after processing every read. In the *batch mode*, reads are processed in batches and updates are made after processing all reads of a batch, which can be an appropriate solution, e.g, for the hybrid approach (see Discussion).

### Compact representation of variant statistics

As opposed to traditional offline consensus callers, an online caller can keep only a very reduced information about alignments. As a result, a space associated with a single position must be limited to a few bytes if we want to keep the statistics in RAM.

For example, in the case of human genome and 6 GB of dedicated RAM (note that the caller is likely to run jointly with other programs such as a read mapper), only up to two bytes per position can be used by the consensus caller. If indels are not considered, the full statistics should contain four integer counters per position, one per nucleotide. Our solution is to keep only most significant bits of each counter. If a counter is saturated but should be incremented, we first bit-shift all the four counters to the right losing the rightmost (least significant) bit, and then increment the counter. An example is given in Fig. 2. This mechanism makes it possible to compute nucleotide frequencies in a limited space and, as a side effect, to filter out sequencing errors that are expected to be randomly distributed. At the beginning of the calling procedure, counters are initialized according to the reference sequence (if it is provided), then counters are updated according to the read nucleotides aligned at the corresponding position.

### Update strategy

When a counter is incremented, we have to decide whether this triggers an update. We propose two different strategies for updates, a *majority strategy* and a *stochastic strategy*. For the sake of simplicity and speed, both strategies consider genomic positions independently and take the decision on the basis of the counters associated with the position. This approach can be potentially improved, at the cost of an additional time, by considering a small window and using Bayesian inference (see, e.g., [25, 33]) to decide on updates.

Let $C_\alpha^{(j)}$ be the value of the $\alpha$-counter at position $j$ for a nucleotide $\alpha \in \{A, C, G, T\}$ and let $C^{(j)} = \sum_{\alpha \in \{A,C,G,T\}} C_\alpha^{(j)}$.

**Majority strategy.** The reference sequence at position $j$ is updated to $\alpha$ once the condition $C_\alpha^{(j)} \geq \frac{1}{2} C^{(j)}$ holds, i.e., $\alpha$ reaches the majority at this position. This strategy results in a moderate number of updates which tend to vanish when the updating process reaches saturation. On the other hand, this strategy may not be flexible enough to avoid the alignment bias.

**Stochastic strategy.** When a counter is incremented, a new base at the corresponding position is drawn randomly from current counter values: the base is set to $\alpha$ with probability $\frac{C_\alpha^{(j)}}{C^{(j)}}$. Note that each sequencing error present in a read causes an increment of the corresponding counter and can, with a small probability, flip the corresponding nucleotide in the reference. Therefore, small fluctuations of the reference sequence can occur even in the hypothetical case of the reference being equal to the sequenced genome. However, a major advantage of this strategy is the reduction of the alignment bias. For diploid or polyploid genomes, the reference will be constantly oscillating between all variants. On the downside, this strategy may generate a large number of updates resulting in a significant overhead.

Note that neither of these two strategies depends on a particular counting mechanism, therefore they can be potentially used with a more complex and expressive statistics.

### Implementation

We developed OCOCO (http://github.com/karel-brinda/ococo), an online consensus caller supporting single-nucleotide updates. OCOCO implements both stochastic and majority update strategies as well as both real-time and batch modes. The software is implemented in C++ and can be integrated into a genomic pipeline or linked directly to a mapper as a library. OCOCO currently supports three counters configurations: 16, 32, and 64 bits per position corresponding to 3, 7, and 15 bits per a single nucleotide counter, respectively. Updates in the reference sequence can be either recorded using "unsorted VCF" or can be directly incorporated in a FASTA file (possibly memory-mapped).

## Simulation of dynamic mapping

As a truly dynamic mapping is hard to implement, we simulate it using a static mapper and emulate dynamic operations by index rebuilding. This section presents details of this simulation.

### Simulation algorithm

Suppose that we have a reference genome $G$ and a set of reads $R = \{r_1, \ldots, r_p\}$ that have to be mapped to $G$. A dynamic mapping can be simulated using a static mapper and an off-line consensus caller by iterating $p$ times the following algorithm. Let $G_i$ be the reference genome obtained after iteration $i$, where $i \in \{1, \ldots, p\}$, and let $G_0 = G$. At iteration $i$, we build an index for the reference $G_{i-1}$, map read $r_i$ to $G_{i-1}$, call consensus from alignments of $r_1, \ldots, r_i$, and build a new reference $G_i$ (see Fig. 3B). The difference with the truly dynamic mapping is that instead of modifying the reference index after mapping a new read, we rebuild it from scratch.

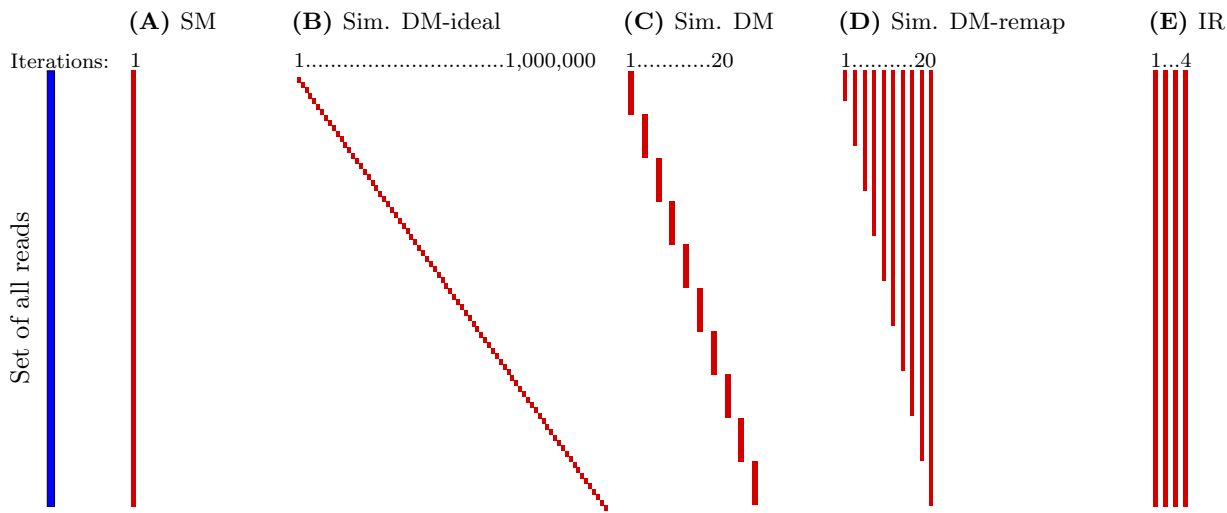

**Figure 3. Simulation of dynamic mapping.** This schematic view shows how dynamic mapping and iterative referencing are simulated in Dynamic Mapping Simulator using a static mapper alone. Subsets of reads to be mapped in individual iterations are shown by vertical lines. In static mapping (SM) (3A), all reads are mapped in a single iteration. Figure 3B illustrates an ideal, but not feasible, simulation when consensus is called after each new read is mapped. In practice, reads are processed by batches (Figure 3C,3D). Dynamic mapping (DM) with (3D) and without (3B,3C) remapping is simulated using nesting or disjoint subsets of reads, respectively. Figure 3E illustrates iterative referencing (IR). In all scenarios 3B - 3E, each iteration is followed by consensus calling and index rebuilding.

One technical point here is that if indel updates are supported by the procedure, the coordinates of downstream loci can be shifted, which would require a transformation of alignments of $r_1, \ldots, r_i$ from the coordinate system of $G_{i-1}$ to that of $G_i$. We will focus on this point in Discussion.

The above approach faithfully simulates dynamic mapping without remapping (i.e. without reconsidering previously computed read alignments), however it cannot be used in practice since the computationally demanding index rebuilding has to be performed too many times. To overcome this obstacle, we observe that at every iteration $i$, new updates can be reported only at positions covered by the newly mapped read $r_i$. We can then partition reads into $t$ batches $Q_1, \ldots, Q_t$, for some $t \ll p$, such that the alignments of any two reads from the same batch do not overlap. This can be achieved, e.g., by assigning only dissimilar reads to the same batch. Now the algorithm works in $t$ iterations and at iteration $i \in \{1, \ldots, t\}$, all the reads from batch $Q_i$ are mapped, which makes the whole procedure fast enough.

In practice, however, computing such batches is a complex task in itself, and instead of fully avoiding overlaps, we minimize them. We choose $t$ large enough and partition reads into $t$ equal-size batches $Q_1, \ldots, Q_t$ (Fig. 3C). If the average coverage per iteration $C_i$ satisfies

$$C_i = \frac{\sum_{r \in Q_i} \text{length(r)}}{\text{length(G)}} \ll 1$$

for every $i \in \{1, \ldots, t\}$, then overlaps are rare.

Until now, we considered dynamic mapping without remapping. To measure the potential effect of remapping, we slightly modify the previous algorithm. At iteration $i$, instead of $Q_i$, we map the whole set $Q_1 \cup \ldots \cup Q_i$ (see Fig. 3D). Note that such an algorithm simulates the ideal situation when a dynamic mapper remaps all reads which can be better aligned after an update of the reference.

To summarize, dynamic mapping is simulated as follows. Given a set of reads and a reference genome $G$, we define $C_i$, compute corresponding number of iterations $t$ and randomly split reads into batches $Q_1, \ldots, Q_t$ of nearly equal size. Then, every iteration consists of three steps: re-building the index according to the current version of the reference, mapping reads, and calling consensus. Depending on whether we choose to perform remapping or not, mapping step $i$ involves $Q_1 \cup \ldots \cup Q_i$ or $Q_i$ respectively.

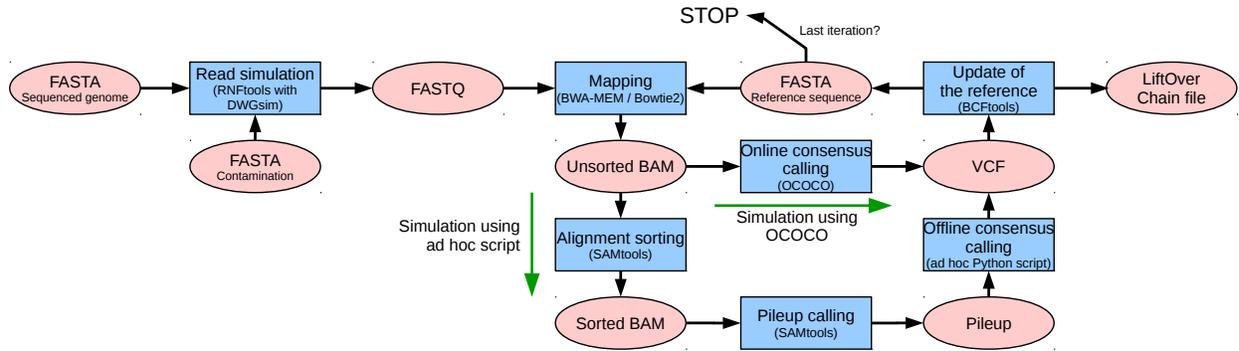

**Figure 4. Dynamic Mapping Simulator: overview of the pipeline.** Reads are selected for mapping as illustrated in Fig. 3. Reference sequence is repeatedly corrected according to the consensus called from already mapped reads. If insertion or deletion updates are allowed, LiftOver Chain file is used to update RNF coordinates in read names.

#### Iterative referencing

To analyze the contribution of dynamic mapping, we compare it to static mapping and, on the other hand, to the so-called *iterative referencing* [36]. Iterative referencing (sometimes called *iterative read mapping* or *iterative remapping*) consists in repeatedly mapping all reads followed by consensus calling, see Fig. 3E. Thus, each update of the reference is inferred from *all* read alignments spanning this locus, and the whole mapping procedure is iterated until the convergence of the reference is reached. Clearly, iterative indexing is a costly method, unfeasible for large datasets, as the index building and the mapping of the whole read set has to be performed several times. On the other hand, iterative indexing tends to produce an optimal variant of the reference that maximizes the overall quality of all alignments. Therefore, in our work we use iterative referencing as a reference for evaluating the quality of dynamic mapping. Various forms of iterative referencing have previously been used in [37–46].

#### Implementation

We developed Dynamic Mapping Simulator (http://github.com/karel-brinda/dymas), a pipeline for comparative evaluation of three mapping methods: static mapping, dynamic mapping (with or without remapping), and iterative referencing (Fig. 4).

For a given reference genome, reads with mutations and sequencing errors are simulated by RNFtools [24] using DWGsim (https://github.com/nh13/DWGSIM). Obtained reads are then distributed into FASTQ files for individual iterations (see Fig. 3). Each iteration starts with a mapping step (using BWA-MEM [5] or Bowtie2 [6]) producing a BAM file. At the consensus calling step, a VCF file with updates is created. Two ways of calling consensus have been implemented. One way is calling consensus using Ococo (see the previous section) directly from unsorted BAM files coming from the mapper. An alternative way is sorting the alignments and producing a pileup file using SAMtools [47], followed by consensus calling via an ad hoc Python script. Finally, reported updates are incorporated into the reference using BCFtools [48] and the index is rebuilt.

Note that consensus calling using Ococo is much faster since it avoids the time-demanding sorting step, however it supports only substition updates. In calling from pileup, deletions and insertions are additionally supported, but the simulation may be slightly less accurate as alignments are processed in the increasing order of starting positions which differs from the order of the original stream. Furthermore, before the evaluation step, read coordinates encoded in RNF names (see below) must be recoded into the current coordinate system in accordance with obtained liftOver chain files.

The entire pipeline is implemented using Snakemake [49] with SMBL (Snakemake Bioinformatics Library, https://github.com/karel-brinda/smbl). Each component (read mapping, pileup computation, consensus calling, etc.) has a standalone easily modifiable Python class. In several auxiliary scripts, GNU Parallel [50] has been used.

| Experiment | Reference genome | Contaminating genome |
|---|---|---|
| Exp. 1 | *Borrelia garinii* (`NC_017717.1`)<br>length 905,534 bp<br>coverage 10× (91,856 reads) | (no contamination) |
| Exp. 2 | *Mycobacterium tuberculosis* (`NC_018143.2`)<br>length 4,411,709 bp<br>coverage 10× (447,481 reads) | *Borrelia garinii* (`NC_017717.1`)<br>length 905,534 bp<br>coverage 5× (45,928 reads) |
| Exp. 3 | *Neisseria meningitidis* (`NC_017513.1`)<br>length 2,184,862 bp<br>coverage 10× (221,616 reads) | *Borrelia garinii* (`NC_017717.1`)<br>length 905,534 bp<br>coverage 5× (45,928 reads) |
| Exp. 4 | *Solibacter usitatus* (`NC_008536.1`)<br>length 9,965,640 bp<br>coverage 10× (1,010,810 reads) | *Borrelia garinii* (`NC_017717.1`)<br>length 905,534 bp<br>coverage 5× (45,928 reads) |

**Table 1. Experiments conducted with Dynamic Mapping Simulator.**

### Evaluation

To compare the three mapping methods (static mapping, dynamic mapping, iterative referencing), we use the RNFTOOLS framework [24]. Within this approach, names of simulated reads store information about their position in the reference. Comparing these coordinates with those reported by mapper allows us to evaluate the correctness of the mapping.

Given a set of BAM files corresponding to iterations of dynamic mapping and iterative referencing, RNFTOOLS creates:

- Data files for all iterations containing the amount of reads of different categories as a function of the threshold on mapping quality.
- Graphs visualizing progressive updates of the ROC curve through individual iterations (see Fig. 5A).
- Graphs displaying fractions of different categories of reads processed in individual iterations (see Fig. 5B).
- Interactive HTML report with all previous data and figures.

All graphs have false discovery rate (FDR) on their $x$-axes. A lower FDR corresponds to a higher threshold on mapping quality, hence a larger number of reads that do not pass quality filtering.

Correctness of an alignment is estimated with respect to its leftmost coordinate. Depending on the tolerance provided to RNFTOOLS, fully-aligned or partially aligned reads can be considered correctly aligned. This allows us to distinguish different consequences of dynamic mapping on the reported alignments.

## Results

**Experimental setup and representation of results.** To evaluate the effect of dynamic mapping, we applied our Dynamic Mapping Simulator in a series of four experiments (Table 1). Each experiment focused on a specific reference genome. In all experiments except one, we considered, in addition, "contamination" reads coming from another genome. This allowed us to analyze the consequences of sample contamination by other genomes as well as presence of reads issued from regions absent in the reference sequence.

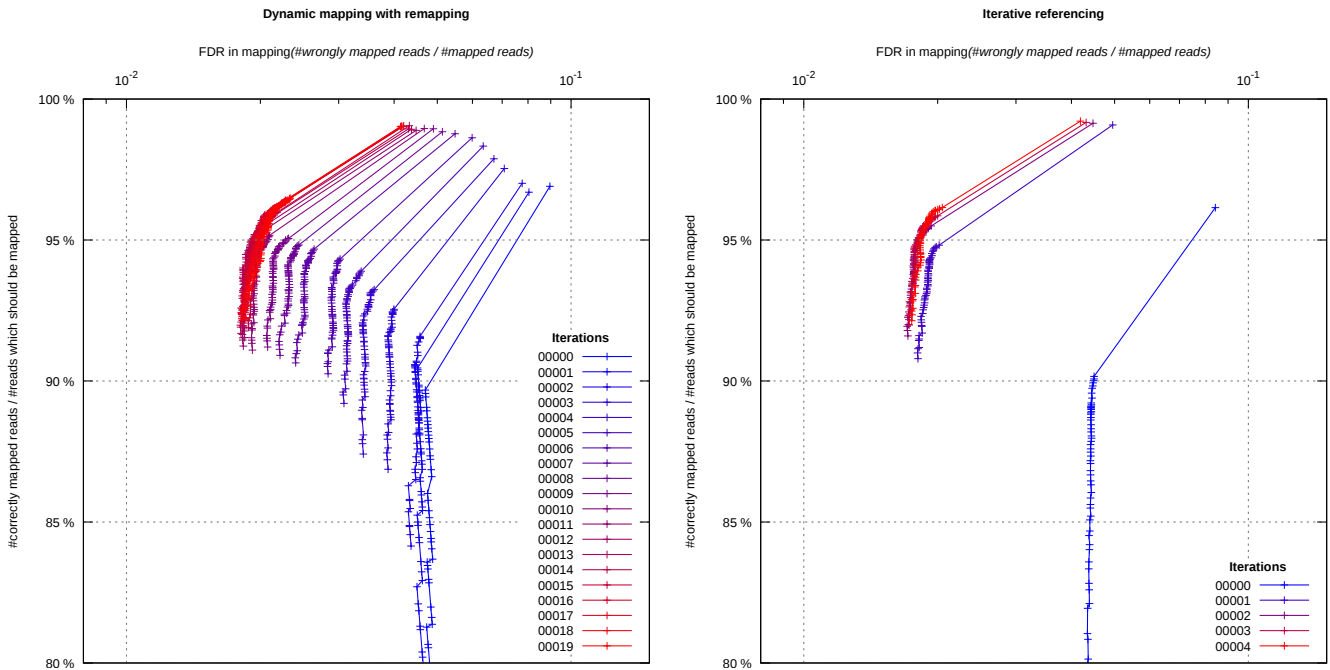

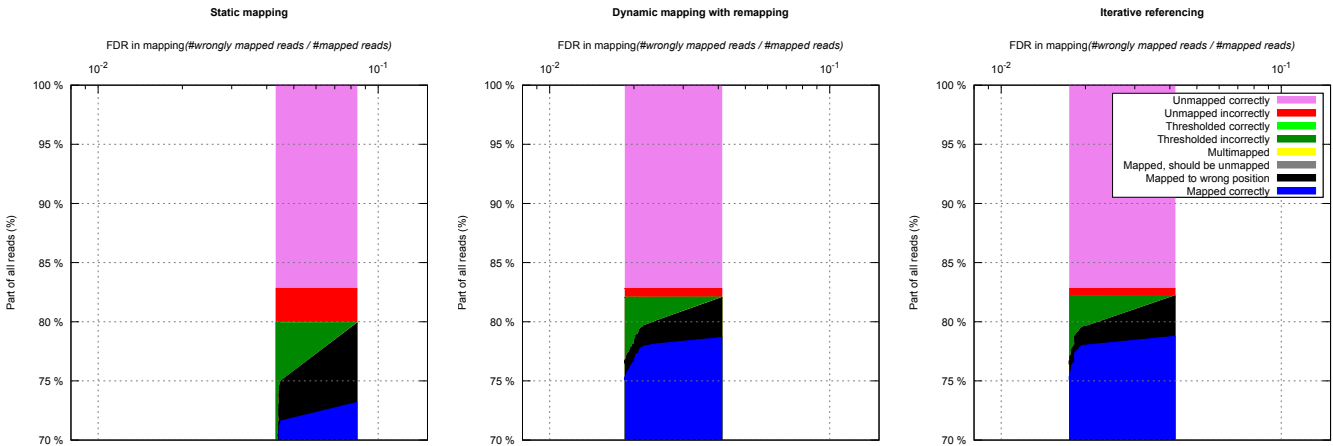

**Figure 5. Detailed report of a selected run of Experiment 3.** Pipelines have been run using BWA-MEM mapping algorithm, with the option of indel calling. Reads were simulated with sequencing error rate 0.02 from mutated *Neisseria meningitidis* genome (mutation rate 0.07) with coverage 10× and contaminated by reads from *Borrelia garinii* with coverage 5×. **(A)** Comparison of ROC (Receiver Operating Characteristic) curves of all iterations of simulated dynamic mapping with remapping (left) and of iterative referencing (right). **(B)** Categories of reads after static mapping, dynamic mapping with remapping, and iterative referencing. Fractions of different categories of reads depending on the level of precision (false discovery rate) for static mapping (left), dynamic mapping with remapping (center) and iterative referencing (right).

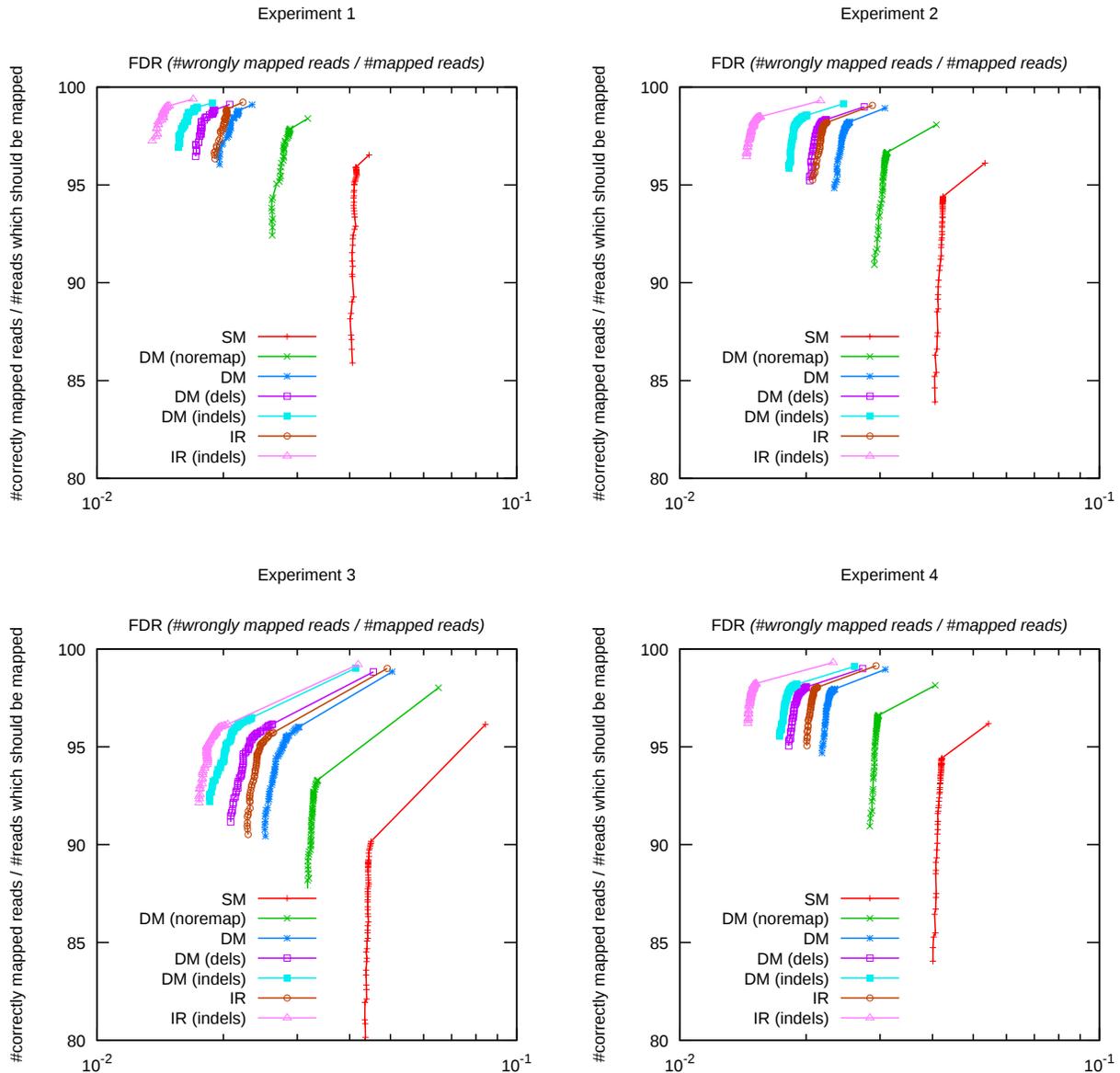

**Figure 6.** Comparison of static mapping (SM), dynamic mapping without calling indels (DM), with calling deletions (DM (dels)), with calling indels (DM (indels)), without remapping (DM (noremap)), and iterative referencing without calling indels (IR), with calling deletions (IR (dels)), with calling indels (IR (indels)) for every experiment.

Experiment 1 evaluates different modes of dynamic mapping on a short genome without any contamination. Experiment 2 highlights the impact of contamination. Experiments 3 and 4 focus on two particular types of bacterial genomes, namely a highly repetitive genome (*Neisseria meningitidis*) and a long genome (*Solibacter usitatus*, 10.0Mbp), respectively.

Within each experiment, we measured the mapping performance for different modes: static mapping, dynamic mapping and iterative referencing with or without remapping, with or without calling deletions, with or without calling both insertions and deletions (indels) (files **S1** and **S2**). The performance of each mapping strategy is represented by a ROC curve on a chart relating the fraction of incorrectly mapped reads (false discovery rate) to the fraction of correctly mapped reads among all reads that should be mapped (Fig. 6 and Fig. 5A). This 'sensitivity-precision' curve uniformly represents the mapping quality on the whole range of parameters and independently of the specific scoring system, score threshold and other involved specific parameters [24]. We also partition reads in eight categories based on whether they are mapped or unmapped correctly (Fig. 5B).

• *Reads that should be mapped* are subdivided into categories *Mapped correctly* (mapped to correct position, passed quality filter), *Mapped to wrong position* (mapped to incorrect position, passed quality filter), *Multimapped* (multiple reported alignments passed quality filter including the correct alignment), *Thresholded incorrectly* (should be mapped but was discarded by quality filter), *Unmapped incorrectly* (incorrect assignment of flag 'unmapped').

• *Reads that should not be mapped* (i.e. coming from contamination) are split into *Unmapped correctly* (correct assignment of flag 'unmapped'), *Mapped, should be unmapped* (mapped and passed quality filter), *Thresholded correctly* (mapped but filtered out).

Technically, in the case of dynamic mapping and iterative referencing, a run of the pipeline produces a set of BAM files, each file storing results of an individual iteration of the mapping algorithm. The final result is considered the one of the last iteration.

**Analysis of the results.** We provide a comprehensive comparative view of different scenarios of dynamic mapping and iterative referencing, contrasted with the regular static mapping (Fig. 6). The first immediate observation is that for each experiment, all curves have a similar shape and can be ordered to have a strictly increasing performance. This supports the observation that their difference is only due to the quality of the reference sequence.

As expected, both dynamic mapping and iterative referencing are significantly better than static mapping. In the first approximation, iterative referencing and dynamic mapping with remapping provide comparable results, with iterative referencing being slightly superior. The latter is natural, as iterative referencing uses all read alignments to make decisions about updates, while dynamic mapping uses only a part of them. On the other hand, results of dynamic mapping without remapping are significantly worse than those of dynamic mapping with remapping. This means that many reads processed in the beginning of the mapping process are inaccurately aligned, which significantly affects the overall results. Thus, correction of those alignments appears an essential step for improving the mapping procedure.

We observe from that indel updates bring a consistent improvement (Fig. 6). Moreover, even supporting deletions alone leads to a more accurate mapping, and accounting for both deletions and insertions of events improves the accuracy even further. Note however that supporting indels by a truly dynamic mapper runs into a hard algorithmic problems of dynamic updates of the underlying data structure and online consensus calling (see Introduction and Discussion).

We found that contamination of read data by sequences from another genome did not have any qualitative consequences on the results. When contamination was introduced, qualities of all runs had been comparably decreased.

**Technical observations.** We also experimented with using BOWTIE 2 as a mapping subroutine. A particularity of BOWTIE 2 is that it has both global and local alignment modes, whereas BWA-MEM performs only local alignment. This allowed us to compare the impact of these modes on the results of the pipeline. Our conclusion is that dynamic mapping should be used in combination with a local alignment algorithm. In the case of global alignment, starting and ending bases of a read tend to be aligned incorrectly in highly mutated regions, which causes statistics corruption and, as a consequence, wrong updates resulting in a damaged reference.

Usage of per-base alignment qualities which has been shown to decrease the amount of false positives in updates [51] has not provided any improvement in our experiments. However, this might be because we do not simulate structural variants for which such a recalibration can improve the accuracy of calling.

We also compared the effect of different amount of statistics stored by a consensus caller (see section above on consensus caller). We found that the default OCOCO option of 16 bits per position did not produce a loss of accuracy compared to the option of 32 bits per position (data not shown).

**Detailed reports.** Detailed information about performed experiments is available within Supporting Information. File **S1** provides a description of parameters of individual runs. File **S2** contains detailed reports similar to Fig. 5 for all runs of every experiment. Full data and HTML reports can be found in directories *Experiments* and *Reports* of `http://github.com/karel-brinda/dymas`.

## Discussion

**Consensus sequence and ploidy.** Even though consensus is primarily used as a technical mean to obtain better alignments, the consensus sequence has a biological significance. If the sequenced genome is haploid, the final consensus tends to correspond to this genome. If it is diploid, the consensus will only contain one variant nucleotide of each SNP, or can even contain none of them if both variants are supported by approximately equal numbers of reads. However, an online consensus calling algorithm can be extended to support diploid genomes as well, provided that ambiguous bases are supported by the mapping algorithm and the underlying index (similarly, e.g., to [52] or [53]). The consensus would then be updated to an ambiguous nucleotide if two distinct variants have been observed at this position.

**Non-substitution updates.** Updates other than substitutions (insertions, deletions, segment reversals, etc.) are difficult to support in dynamic mapping for two main reasons. Firstly, they entail changes in genomic coordinates, and therefore either all reported alignments have to be translated into the coordinate system of the final consensus sequence (which may require local realignment of reads), or differences between the original reference sequence and the current consensus have to be continuously recorded in a dedicated data structure allowing the retrieval of original coordinates. Secondly, appropriate counters and calling algorithms have to be designed for each type of updates.

If we are limited to deletions only, these problems can be solved by using a padded reference sequence (see, e.g., [54]) and introducing specific counters for deletions. Then, deletions are treated in consensus calling in the same way as mismatches, after extending the nucleotide alphabet by an additional letter '∗'. If both insertions and deletions have to be supported, the algorithm becomes much more difficult to implement. Using a padded reference allows one to only support insertions at positions from a pre-specified list (by introducing '∗' at these positions), or at positions where a deletion has previously been made. A full support of insertions of arbitrary length and occurring at arbitrary positions remains an open problem that is still awaiting a practical algorithmic solution.

**Databases of SNPs.** SNP databases such as DBSNP [55] can be incorporated into dynamic mapping in two ways. We can either use an SNP database to set initial values of online consensus caller counters, or restrict the updates only to SNPs recorded in the database. Restricting possible updates would reduce the computational complexity of the whole algorithm and would improve the accuracy of processing those updates by keeping a richer variant statistics and improving indel calling. Note that this approach can be viewed as a particular case of *path projection*, see section Reference graphs.

**Pileup information.** As a side product of dynamic mapping, a simplified pileup information can be obtained (for more information about the pileup format, see SAMTOOLS documentation [47]). Therefore, NGS methods supporting pileup as input format (such as those of [56, 57]) can be combined with dynamic mapping in order to avoid expensive steps of alignment sorting and pileup calling. Note that while dynamic mapping causes a constant-factor slow-down of the mapping step – from time $t$ to time $kt$, where $k$ is a deceleration constant specific to each dynamic mapping algorithm – the sorting step has $n \log(n)$ time complexity, where $n$ is the number of reads, which often makes this step the most time demanding of the pipeline.

In the OCOCO online consensus calling algorithm (see Methods), several pieces of information are lost compared to the "standard pileup" (as produced by SAMTOOLS): information about base qualities and alignment qualities, order of bases (only counts are kept), distinction of strands (forward or reverse), deletions and insertions. Moreover, the pileup is truncated when the counter capacity is lower than the maximal coverage in the experiment.

At the price of an additional memory consumption, the counting mechanism can be modified to recover the distinction of strands (via doubling all counters) and the count of deletions (via adding a specific counter for deletions, see section Non-substitution updates). Truncating can be avoided in Ococo already now by setting the size of counters large enough.

**Hybrid approach.** Many computational pipelines are deployed on clusters of many nodes, which enables to combine online consensus calling with static mapping in a similar way as dynamic mapping without remapping has been simulated. More specifically, one particular node would be assigned to collect alignments from mapping nodes and to call consensus online in the batch mode. Another node, in charge of maintaining the index, could iteratively retrieve the current consensus, rebuild the index and distribute it to mapping nodes. Such a hybrid approach (combining dynamic mapping and iterative referencing) remains to be verified in practice.

**Sequencing in streaming mode.** Most of existing sequencers are technically possible to be run in a streaming mode, and several studies [58–63] already considered the streaming mode in various tasks. Moreover, practical software solutions for streaming sequencing have already been proposed [64].

In the context of dynamic mapping, streaming sequencing can be particularly beneficial, as produced reads can be directly piped to a dynamic mapper. Then the sequencing can be kept on until the internal statistics gets saturated and the consensus sequence becomes stable enough. After these combined sequencing and mapping steps, only a certain amount of last reported alignments should be used for the subsequent analysis. The reason is that, as follows from the comparison of dynamic mapping with and without remapping (Fig. 6), reads mapped to a mature consensus are aligned more accurately.

**Detection of somatic cancer mutations.** Somatic cancer mutations are often very underrepresented in the reads as tumor samples are contaminated by normal cells and, on the other hand, somatic mutations are often specific to particular subclones of the tumor. Therefore, frequencies of somatic cancer mutations can be even lower than the sequencing error rate, which makes their accurate detection very hard (see, e.g., [65]). To deal with this issue, normal (control) samples need to be sequenced in addition to tumor samples to allow for a comparative analysis – see, e.g., papers [56, 65–71] on detecting single-nucleotide variants or papers [56, 57, 72, 73] on copy-number abberations.

To deal with low mutation frequencies, methods have to rely on high-quality alignments. As we showed in our work, dynamic mapping can produce significantly more accurate alignments, which results in a better identification of somatic mutations proximal to SNPs. Note that cancer SNV databases such as COSMIC [74] can be incorporated into dynamic mapping, as it was with SNP databases (see section Databases of SNPs).

Low somatic mutation frequencies make it necessary to employ a high sequencing coverage which in turn results in a longer processing time because of the slow alignment sorting step. However, this step can be avoided by working directly with the simplified pileup (see section on Pileup information).

**Long reads.** Sequencing technologies producing long reads, such as *Pacific Biosciences single-molecule real-time sequencing* or *Oxford Nanopore sequencing*, have recently emerged. Compared to traditional NGS platforms such as *Illumina MiSeq* or *HiSeq*, they produce considerably longer reads with higher rates of errors, especially indels. This situation calls for new algorithms specifically designed for long reads, and dynamic mapping can be helpful in several ways here.

Some methods of long read correction (see, e.g., [75]) proceed by aligning short reads to long reads, possibly in a way similar to iterative referencing (see, e.g., [44]). Within these alignment-based methods, dynamic mappers have a great potential to improve the mapping step. However, considering the type of errors in long reads, indel updates have to be supported by the mapper (see section Non-substitution updates).

Another perspective is applying dynamic mapping of long reads. Unfortunately, consensus calling from long read alignments appears to be a hard problem even in the offline setting, due to observed error profiles. For instance, genome assembly algorithms apply sophisticated techniques based on partial order graphs to obtain a good consensus (see, e.g., [76]). In order to adapt dynamic mapping to long reads, update strategies in online consensus calling have to be significantly improved, in particular a single update decision has to take into account counter information at multiple positions. This approach requires a better statistical modeling and deserves further research.

**Guided genome assembly.** If a high quality reference genome is not available, the approach based on read mapping and variant calling cannot be applied directly. Instead, a computationally expensive genome assembly (see, e.g., [77]) becomes unavoidable. Nevertheless, when a reference genome for a related species is available, it can be exploited in order to speed up the assembly process or to improve the quality of produced assemblies. This approach is known as *reference-guided assembly* or *assisted assembly*. Either a single [78–80] or multiple [81] reference genomes are used as references for auxiliary read alignment to create contigs, or a reference helps to combine assembled contigs [82–85]. In those approaches where mapping *of* or *to* contigs is employed, dynamic mapping can help in a similar way to what was discussed in section Long reads (contigs can be viewed as a special case of long reads).

**Phylogenetic inference.** A class of methods of phylogenetic inference relies on read mapping. In those methods, reads are mapped to one or multiple references, SNP positions are extracted and a phylogenetic tree is reconstructed, typically using a maximum likelihood technique. It has been shown [86] that even in the case of relatively low degree of divergence between queried and reference sequences, the mapping bias can cause inaccurate estimations of the distance which results in incorrect tree topologies. The correction capacity of dynamic mapping could reduce this bias and produce better phylogenetic distances, which constitutes a promising application of dynamic mapping.

**Reference graphs.** As the traditional concept of sequences as reference structures has become insufficient [87], *reference graphs* (sometimes also called *variation graphs*) appear to be a more appropriate model than *reference sequences*. Therefore, new techniques of mapping to reference graphs [88–95], as well as implementations of reference graphs [96–98] are now a subject of intensive research.

Since state-of-the-art graph mappers (such as VG (https://github.com/vgteam/vg) with GCSA2 [95] as indexing component) are still experimental and no dynamic mapper has been developed so far, a hypothetical *dynamic graph mapper* can now be studied only from a theoretical viewpoint. Two ways of combining dynamic mapping and mapping to reference graphs can be considered: *path projection* and *dynamic graphs*.

In the *path projection* approach, a fixed graph reference is projected onto a linear sequence. According to observed variants, the reference sequence is updated in such a way that it still corresponds to some path in the graph reference. Therefore, this can be considered as a form of dynamic mapping with a restricted set of allowed updates specified by the graph reference. Decisions about updates can be done stochastically, similar to stochastic consensus calling described in the Methods section. In such an approach, the probability of a path to be selected should reflect the fraction of reads already mapped to this path and their mapping qualities.

Implementing path projection requires a dynamic mapper with a particular online consensus caller working with a graph reference. Instead of collecting statistics for individual bases, it should collect statistics for paths in the graph. The main advantage of this approach is that it can be built on top of a mapping algorithm working with linear references (such as BWA or BOWTIE2). Note that very dissimilar parallel paths in the graph could be decomposed into several smaller graphs to prevent extensive updates of the linearized sequence.

In the *dynamic graphs* approach, the reference graph itself is updated according to reads mapped so far. The graph is either extended by incorporating all observed variants as new paths, or modified within the same topology, or both. In the first case, online consensus calling can be completely avoided if *all* observed differences are incorporated to the graph. However, all sequencing errors would then get included, which might result in a massive growth of the graph. In the second case, an online consensus caller could be similar to OCOCO but with a support for reference graphs. In both cases, the underlying graph structure and the corresponding index must support updates.

**Future work.** Implementing an efficient dynamic mapper remains an open problem, however most critical ingredients of such an implementation have already been developed. An online consensus calling algorithm has been implemented in the OCOCO program within our work, and a promising dynamic $k$-mer index called SKIPPATCH has been recently reported in [21]. Altogether, implementing a dynamic mapper limited to substitution updates and "restricted" indel updates (see Non-substitution updates in Discussion) appears nowadays as a challenging but mainly engineering task.

Several other algorithmic problems remain open when more general updates are considered. The main open problem is to design a more complex framework to maintain variation statistics in compact space through the mapping procedure, including information about arbitrary-length insertions and large-scale variants (inversions, transpositions, etc.). A related problem is to define an optimal strategy for reference updates, based on the collected statistics. This may involve building an appropriate probabilistic model, generalizing the stochastic strategy we considered in this work.

With a fully dynamic mapper available, the improvement of alignment quality will have to be evaluated for different tasks, such as variant calling, somatic mutation analysis, reference guided assembly, phylogenetic inference, or processing long reads. Several new perspectives of NGS data processing will open up, such as real-time mapping or real-time reference-based variant calling.

It would be also interesting to explore applications of online consensus calling other than dynamic mapping, e.g. to a distributed implementation of mapping with specific processor nodes dedicated to correcting the reference and rebuilding the index (see Hybrid approach).

## Conclusions

We have shown that the dynamic mapping approach can notably improve read alignment and variant calling. This improvement is especially important when the sequenced genome is highly mutated compared to the reference, or belongs to a different species. Thus, in numerous applications involving highly mutated regions (e.g., hotspot recombination regions), dynamic mapping can be beneficial. We showed that dynamic mapping can potentially reach an accuracy comparable to that of iterative referencing, provided that remapping of reads is enabled. Note that iterative referencing approximates the optimal mapping of all reads to a modified reference. On the other hand, we found that if remapping is not performed, the accuracy of dynamic mapping becomes significantly lower, but is still notably better than that of the regular static mapping.

As other results of our work, we provided software programs for simulating dynamic mapping: Dynamic Mapping Simulator (`http://github.com/karel-brinda/dymas`) and Ococo online consensus caller (`http://github.com/karel-brinda/ococo`). To our knowledge, Ococo is the first consensus caller supporting the online mode, however its accuracy is limited by restricting possible variation to substitutions only.

On the other hand, our conclusion is that a fully dynamic mapper should be significantly slower than an optimized static mapper, due to algorithmic difficulties that have to be overcome. This implies that practical benefit of dynamic mapping will be mainly in those scenarios where iterative re-building of the index cannot be afforded (e.g., because of a large size of the reference sequence).

## Supporting Information

**S1 File. Description of runs in a single experiment.**

**S2 File. Detailed reports for all runs of all experiments.**

## Acknowledgments

KB and GK were supported by the ABS4NGS grant and by Labex Bézout of the French government (program Investissement d'Avenir). VB was supported by the ATIP-Avenir Program, the ARC Foundation and the "Who Am I?" Project.